\documentclass[a4paper,11pt]{article}

\usepackage{a4wide}
\usepackage{amsmath}
\usepackage{amssymb}
\usepackage[caption=false,font=footnotesize]{subfig}

\usepackage{tikz}
\usetikzlibrary{decorations.pathmorphing}
\usetikzlibrary{shapes.symbols}
\usetikzlibrary{chains} 
\usetikzlibrary{positioning} 
\usetikzlibrary{calc} 
\usetikzlibrary{decorations.pathreplacing} 
\tikzset{
  vertex/.style = {
    circle,draw, fill = white, inner sep = 0pt, minimum width = 4pt  }
}

\usepackage{algorithm}
\usepackage{algorithmic}
\newtheorem{thm}{Theorem}
\newtheorem{lem}[thm]{Lemma}

\newtheorem{cor}[thm]{Corollary}

\newenvironment{pf}[1][Proof]{\begin{trivlist}
\item[\hskip \labelsep {\bfseries #1}]}{\end{trivlist}}

\newcommand{\qed}{\nobreak \ifvmode \relax \else
      \ifdim\lastskip<1.5em \hskip-\lastskip
      \hskip1.5em plus0em minus0.5em \fi \nobreak
      \vrule height0.75em width0.5em depth0.25em\fi}

\begin{document}

\newcommand{\name}[1]{\textsc{#1}}

\newcommand{\wit}{\odot}
\newcommand{\bmm}{\cdot}
\title{Counting Perfect Matchings as Fast as Ryser\thanks{This work was supported by the Swedish Research Council grant ``Exact Algorithms'' VR 2007-6595.
}}

\author{Andreas Bj\"orklund\thanks{Dept. of Comp. Sci., Lund University.}}

\date{}

\maketitle 

\begin{abstract}
We show that there is a polynomial space algorithm that counts the number of perfect matchings in an $n$-vertex graph in $O^*(2^{n/2})\subset O(1.415^n)$ time. ($O^*(f(n))$ suppresses functions polylogarithmic in $f(n)$).The previously fastest algorithms for the problem was the exponential space $O^*(((1+\sqrt{5})/2)^n) \subset O(1.619^n)$ time algorithm by Koivisto, and for polynomial space, the $O(1.942^n)$ time algorithm by Nederlof.
Our new algorithm's runtime matches up to polynomial factors that of Ryser's 1963 algorithm for bipartite graphs. We present our algorithm in the more general setting of computing the hafnian over an arbitrary ring, analogously to Ryser's algorithm for permanent computation.

We also give a simple argument why the general exact set cover counting problem over a slightly superpolynomial sized family of subsets of an $n$ element ground set cannot be solved in $O^*(2^{(1-\epsilon_1)n})$ time for any $\epsilon_1>0$ unless there are $O^*(2^{(1-\epsilon_2)n})$ time algorithms for computing an $n\times n$ $0/1$ matrix permanent, for some $\epsilon_2>0$ depending only on $\epsilon_1$.
\end{abstract}

\section{Introduction}
The permanent of a  matrix $A=(A_{i,j})\in R^{n\times n}$ over a ring $R$ is defined as
\[
\mbox{per(A)}=\sum_{\sigma\in S_n} \prod_{i=1}^n A_{i,\sigma(i)}
\]
Here $S_n$ is the set of all permutations on $n$ elements.
Ryser \cite{R63} gave an algorithm for computing the permanent in $O(n2^n)$ operations over $R$, and it remains to this date the fastest known for general matrices. Only in special cases as e.g.  when the input matrix is sparse significantly faster algorithms are known. In this paper we show that a previously studied, seemingly harder, problem admits just as fast an algorithm up to polynomial factors. It includes the permanent computation as a special case:
The hafnian of a matrix $B=(B_{i,j})\in R^{2n\times 2n}$ over a ring $R$ is commonly defined as
\[
\mbox{haf}(B)=\sum_{\sigma\in C_{2n}} \prod_{i=1}^n B_{\sigma(2i-1),\sigma(2i)}
\]
Here $C_{2n}$ is the set of canonical permutations on $2n$ elements, permutations $\sigma$ obeying I. $\forall i:\sigma(2i-1)<\sigma(2i)$ and II. $\forall i:\sigma(2i-1)<\sigma(2i+1)$. Observe that hafnians only depend on matrix elements strictly above the main diagonal. For conventional reasons though, a hafnian is defined only for zero diagonal symmetric matrices. A simple connection between hafnians and permanents is given by
\[
\mbox{haf}\left(\left[\begin{array}{ll} 0 & A\\ A^{\small T} & 0 \end{array} \right]\right)=\mbox{per}(A)
\]


We show that $2n\times 2n$ hafnians can be computed as fast as Ryser's algorithm computes $n\times n$ permanents.
\begin{thm}
\label{thm: haf}
For $B\in R^{2n\times 2n}$ a symmetric zero diagonal matrix with elements from a ring $R$, $\operatorname{haf}(B)$ can be computed in $O^*(2^n)$ ring operations, storing only a polynomial in $n$ number of ring elements at any instance of the computation.
\end{thm}

Computing the permanent of an $n\times n$  $0/1$ matrix has an equivalent formulation in terms of counting perfect matchings in an associated bipartite graph on $2n$ vertices. Just identify rows with one half of the vertices and the columns with the other. There is an edge between row vertex $r$ and column vertex $c$ if the matrix entry at $r,c$ is $1$. Analogously, computing the hafnian of a $2n\times 2n$  $0/1$ matrix has an equivalent formulation in terms of counting perfect matchings in an associated \emph{general} graph on $2n$ vertices. Think of the matrix as an adjacency matrix for a graph.
Thus with our new algorithm we can count perfect matchings in general graphs just as fast as Ryser's algorithm counts them in bipartite ones. 
\begin{cor}
\label{cor: pm}
There is a polynomial space, $O^*(2^{n/2})$ time algorithm that computes the number of perfect matchings in any $n$-vertex graph.
\end{cor}

In other words, either bipartiteness doesn't help much for perfect matching counting, or Ryser's algorithm is far from optimal. This should be contrasted with the problems' known approximability situation: There are polynomial time approximation schemes for counting perfect matchings in bipartite graphs \cite{JSV04}, but the fastest approximation schemes for general ones still run in time singly exponential in the number of vertices \cite{S03}.

We note that the technique used in our algorithm makes profound use of the fact that it counts covers of $2$-sets. One would hope that the technique could bear fruit even for larger set families. Unfortunately it seems like it doesn't. We give here some evidence to why this may be. In a general exact set cover counting problem, we are given
a family $\mathcal{F}$ of subsets of arbitrary size of a ground set $U$ on $n$ elements. The objective is to count the number of ways to pick pairwise disjoint subsets from $\mathcal{F}$ whose union cover all of $U$. This is known to admit an $O^*(2^n)$ time algorithm \cite{BHK09} (and a trivial $O^*(|\mathcal{F}|2^n)$ time algorithm by dynamic programming across all vertex subsets). We prove here that if the fastest known permanent computation algorithms for $0/1$ matrices are close to optimal, then we cannot hope to improve on general exact set cover counting.

\begin{thm}
\label{thm: hard}
For every constant $\epsilon_1>0$, any algorithm capable of counting the solutions to any exact set cover instance $\mathcal{F},U,|U|=n$, $|\mathcal{F}|\in n^{O(\log n)}$ in $O^*(2^{(1-\epsilon_1)n})$ time, implies the existence of an algorithm that computes the permanent of any $n\times n$ $0/1$ matrix in $O^*(2^{(1-\epsilon_2)n})$ time, for some $\epsilon_2>0$ depending only on $\epsilon_1$.
\end{thm}

\subsection{Previous Work}
\subsubsection*{Computing the Permanent.}
Ryser~\cite{R63} presented an inclusion--exclusion based algorithm to evaluate the permanent of an $n\times n$ matrix over any ring in $O(n2^n)$ ring operations.
Valiant~\cite{V79} showed that computing the permanent even restricted to elements of non-negative integers is $\#\mbox{P}$-hard. Dell et al.~\cite{DHW10}  concluded in the same spirit that one cannot compute $n\times n$ integer permanents in $2^{o(n)}$ time unless you can count the satisfiying assignments to 3CNF Boolean $n$-variate formulas faster than $\Omega(c^n)$ for every $c>1$ (I.e. a counting analogue of the Exponential Time Hypothesis \cite{IPZ01}). There is however no known explanation as to why the exponential base $2$ should be optimal. 

For restricted cases, there are some small improvements.
Bax and Franklin~\cite{BF02} showed that for $0/1$ matrices, there is an $exp[-½ ( n ^{1/3}/2 ln(n) )]2^n$ expected time algorithm. Building on their construction, Servedio and Wan \cite{SW05} gave an algorithm that computes the permanent of any matrix with at most $cn$ nonzero entries in $O^*((2-\epsilon)^n)$ time, with $\epsilon$ depending on $c$. 

\subsubsection*{Counting Perfect Matchings in General Graphs.}
Bj\"orklund and Husfeldt~\cite{BH08} gave a polynomial space algorithm that counts the perfect matchings in an $n$-vertex graph in $O(n^22^n)$ time. They also presented an exponential space $O(1.733^n)$ time algorithm based on fast matrix multiplication. Koivisto~\cite{K09} showed an elegant exponential space $O^*(\phi^n)$ time
algorithm where $\phi=(1+\sqrt{5})/2\approx 1.618$ is the golden ratio. 
A generalization of Ryser's formula to general graphs was given independently by Nederlof\cite{N08}, and Amini et al. \cite{AFS09}, who showed that if the graph has an independent set of size $i$, one can count the perfect matchings in $O(n^32^{n-i})$ time. In a yet unpublished paper Nederlof \cite{N10} further improves the polynomial space running time to $O(1.942^n)$.
  
\subsection{Overview.}
The remainder of the paper is organized as follows.
In Section~\ref{sec: idea} we try to explain the idea behind the algorithm. A formal proof in the hafnian setting is given in Section~\ref{sec: haf}, where the proofs of Theorem~\ref{thm: haf} and indirectly Corollary~\ref{cor: pm} are given. Finally, In Section~\ref{sec: hard} we give the proof of Theorem~\ref{thm: hard}.

\section{Our Algorithmic Idea}
\label{sec: idea}
\begin{figure}[htb]
\[
\begin{tikzpicture}
    \draw 
  	(-1,2) node (b) [vertex,label=below left:\small$b$] {}
	(-1,4) node (a) [vertex,label=above left:\small$a$] {} 
  	(0,4) node (c) [vertex,label=above :\small$c$] {}
	(0,2) node (d) [vertex,label=below :\small$d$] {}
     	(1,4) node (e) [vertex,label=above right:\small$e$] {}
	(1,2) node (f) [vertex,label=below right:\small$f$] {};
     
   \begin{scope}[very thick]
     \draw  (a)--(b) (a)--(c) (a)--(d) (a)--(f) (b)--(c) (b)--(d) (c)--(d) (c)--(e) (e)--(f) (f)--(d);
    \end{scope}
   
   \draw (0,0) node {Stage $0$};
\end{tikzpicture}
\qquad
\begin{tikzpicture}
 \qquad
  \draw (0,1) node {{\small $\emptyset: ef[1].$}};
  \draw (1.55,3.0) node {{\small $ce\cup df[1]$}};
 \draw (0,4.7) node {{\small $af\cup ce[1]$}};
   \draw 
  	(-0.5,2) node (b) [vertex,label=below left:\small$b$] {}
	(-0.5,4) node (a) [vertex,label=above left:\small$a$] {} 
  	(0.5,4) node (c) [vertex,label=above right:\small$c$] {}
	(0.5,2) node (d) [vertex,label=below right:\small$d$] {};
	
 \begin{scope}[very thick]
	\draw (a)--(b) (a)--(c) (a)--(d) (b)--(c) (b)--(d) (c)--(d) ;
\end{scope}
\begin{scope}

  \draw  (c) .. controls (0.5,3) .. (d);
   \draw  (a) .. controls (0,4.2) .. (c);

\end{scope}
\draw (0,0) node {Stage $1$};
\end{tikzpicture}
\qquad
\begin{tikzpicture}
  \draw (0,1) node {{\small $\emptyset: ef[1],cd[2],$}};
  \draw (0,0.5) node {{\small $ce \cup df[12].$}};
  
  \draw (1.55,2.5) node {{\small $af\cup ce\cup bd[12].$}};
  \draw (1.15,3.0) node {{\small $ac\cup bd[2],$}};
  \draw (1.1,3.5) node {{\small $ad\cup bc[2],$}};   

   \draw 
  	(0,2) node (b) [vertex,label=below :\small$b$] {}
	(0,4) node (a) [vertex,label=above :\small$a$] {};
	
 \begin{scope}[very thick]
	\draw (a)--(b);
\end{scope}
\begin{scope}

  \draw  (a) .. controls (0.3,3) .. (b);

\end{scope}
\draw (0,0) node {Stage $2$};
\end{tikzpicture}
\qquad
\begin{tikzpicture}

\draw (0.25,4.5) node {{\small $\emptyset:$}};
  \draw (0.9,4) node {{\small $ef[1],$}};
  \draw (0.9,3.5) node {{\small $cd[2],$}};
  \draw (0.45,3) node {{\small $ce \cup df[12],$}};
  \draw (0.9,2.5) node {{\small $ab[3],$}};
  \draw (0.45,2) node {{\small $ad\cup bc[23],$}};
  \draw (0.45,1.5) node {{\small $ac\cup bd[23],$}};
  \draw (0,1) node {{\small $af\cup ce \cup bd[123].$}};

  \draw (0,0) node {Stage $3$};

\end{tikzpicture}
\]

\caption{ 
\label{fig}
 Vertices are removed in pairs and are replaced by labels, first vertices $e,f$ for the label $1$, then $c,d$ for $2$, and finally $a,b$ for $3$. Labeled edges including empty edges listed under ``$\emptyset:$'', representing partial matchings of the removed vertices, are created. The labels are written within brackets. In the final stage, the partial matchings listed under ``$\emptyset:$'' constitute an exact set cover problem for which the solutions are precisely the perfect matchings in the original graph.
}
\end{figure}
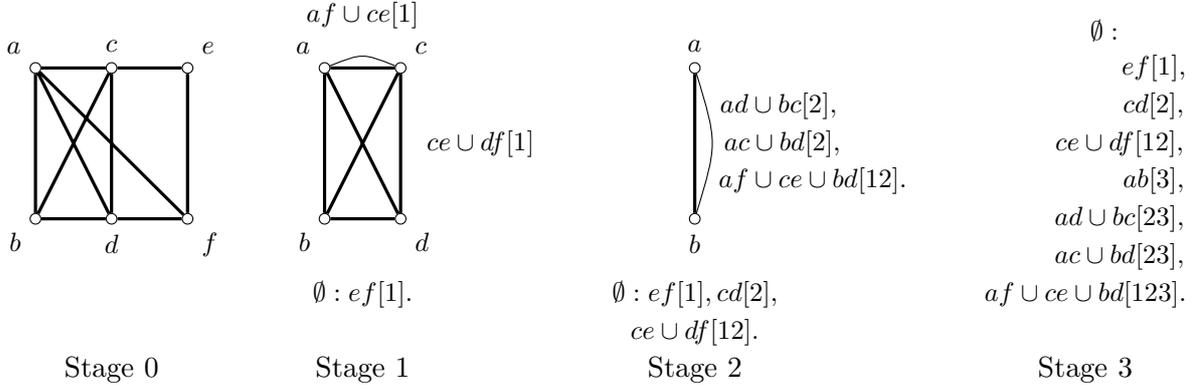

We will in the next section describe our new algorithm in detail. First we will try to provide some intuition. The underlying idea of the algorithm is perhaps best envisioned in the perfect matching setting. Consider an undirected unweighted graph and suppose you want to count its perfect matchings. Our algorithm operates in stages. In each stage we remove two vertices from the graph and replace them with new ``labeled'' edges between the remaining vertices. The added labeled edges represent ways a partial matching can cover the removed vertices. In particular, the partial matching always cover both of the removed vertices and thus they can be treated as just one label, we never consider a partial matching covering only one of them. The effective speedup is achieved exactly by this fact, that we always trade two vertices for one new label. Confer Figure~\ref{fig} for an example of three subsequent invocations of this remove-and-replace operation. As the operation is repeated it results intermediately in multilabeled edges. When the graph's vertices are exhausted, all that is left is a set of multilabeled ``empty'' edges. Their disjoint label covering combinations describe all perfect matchings of the original graph. This is an exact set cover counting problem on half of the vertices number of labels. In Figure~\ref{fig}, all ways to combine the partial matchings listed under ``$\emptyset:$'' in stage $3$ such that each of the three labels $1,2,3$ is used precisely once, represents a perfect matching in the original graph in stage $0$. E.g. $ef[1],ad\cup bc[23]$ is the  perfect matching $ef\cup ad\cup bc$, and $ce\cup df[12],ab[3]$ is the perfect matching $ce\cup df \cup ab$. Moreover, every perfect matching in the original graph can be written as an exact set cover from the list in this way.
In the next section we will see how this idea is formalized in the general hafnian setting.

\section{An Algorithm for the Hafnian}
\label{sec: haf}
In this section we prove Theorem~\ref{thm: haf}. We describe an algorithm to compute $\mbox{haf}(B)$ of a matrix $B\in R^{2n\times 2n}$ for an arbitrary ring $R$.  First we suggest an implementation of our algorithm that uses exponential space since it is easier to reason about, but we will see  in Section~\ref{sec: polyspace} how the space usage can be reduced to only polynomial in $n$.

\subsection{Label Extensions of a Ring.}
We will operate over extension rings of $R$.  They are designed to keep track of the multilabeled elements mentioned in the previous section. The reason will hopefully become clear in Section~\ref{sec: alg} where we prove a syntactically tight recursion formula for the hafnian in terms of the extension ring. We call the extension rings $R[U_m]$ for $m$ a non-negative integer.
One can think of the structure $U_m$ as the set of subsets of $[m]$, with a partial function $\cup_{disj}:[m]\times [m]\rightarrow [m]$ defined as $\cup_{disj}(a,b)=a\cup b$ for disjoint $a,b$ and undefined otherwise.

An element in $R[U_m]$ is nothing more than a vector $r\in R^{[m]}$, i.e. a vector of $2^m$ elements from $R$. For an element $r\in R[U_m]$, we write $r_X$ to denote the coordinate in $r$ associated with the subset $X$. Syntactically, we write 
\[
r=\sum_{X\subseteq [m]} r_X[X]
\]
to describe the ring extension element $r$.
Addition of $r,q\in R[U_m]$ is given by 
\[
r+q=\sum_{X\subseteq [m]} (r_X+q_X)[X]
\]
Here the $+$ within the parantheses refers to addition of elements in $R$, i.e. the vectors $r$ and $q$ are added elementwise.
Multiplication is a bit more complicated. It is defined in a convolution like manner by 
\[
r\cdot q=\sum_{X\subseteq [m]} (\sum_{Y\subseteq X} r_Yq_{X-Y})[X]
\]
Again, operations within the parantheses are over the ring $R$. We write $1[\emptyset]$ to emphasize the multiplicative identity. Note that $R[U_i]$ is a subring of $R[U_{i+1}]$ for every $i$, i.e. operations over $R[U_i]$ are meaningful also in $R[U_{i+1}]$. We will use this fact extensively.

\subsection{Algorithm.}
Suppose we want to compute $\mbox{haf}(B)$ for $B\in R^{2n\times 2n}$.  
Our algorithm operates in stages. Initially, we set $B^{(0)}=B$. In each subsequent stage $i$, we will reduce the hafnian computation of a matrix $B^{(i-1)}$ to one for a matrix $B^{(i)}$. We call this operation a \emph{squeeze}, because the latter matrix's dimensions will be smaller, but its elements belong to a larger ring than its predecessor's.
Formally, in stage $i$ the hafnian of a matrix $B^{(i-1)}\in R[U_{i-1}]^{(2n-2i+2)\times (2n-2i+2)}$ is related to the hafnian of a derived matrix $B^{(i)}\in R[U_{i}]^{(2n-2i)\times (2n-2i)}$. In the final stage $n$, we are left with a hafnian of a zero-dimensional matrix which we by definition equate with one. We also associate an element $\beta^{(i)}\in R[U_{i}]$ with stage $i$ called the \emph{squeeze factor} of the stage. Our algorithm simply computes the product of these squeeze factors and in the end returns the element associated with the label set $[n]$, confer Algorithm~\ref{alg: haf}.

\begin{algorithm}
\caption{\label{alg: haf}
haf($B$),$\,B\in R^{2n\times 2n}$}
\begin{algorithmic}
\STATE $B^{(0)}\leftarrow B$
\STATE $h\leftarrow 1[\emptyset]$
\FOR{$i=1,2,\cdots n$ }
\STATE   Compute $B^{(i)}$ and $\beta^{(i)}$ (according to Section~\ref{sec: squeeze}).
\STATE $h\leftarrow h\cdot \beta^{(i)}$
\ENDFOR
\RETURN $h_{\left[n\right]}$
\end{algorithmic}
\end{algorithm}

\subsubsection{The Squeeze Operation.}
\label{sec: squeeze}

In every stage we will expunge the two first rows and columns from the present input matrix and instead append a new element (a label) to the set of the ring extension. 
Consider stage $i$. 
On input $B^{(i-1)}\in R[U_{i-1}]^{(2n-2i+2)\times (2n-2i+2)}$ we define $B^{(i)}\in R[U_{i}]^{(2n-2i)\times (2n-2i)}$ through
\[
B^{(i)}_{j,k}=
\left\{\begin{array}{ll} B^{(i-1)}_{j+2,k+2}+ S_{i,j,k} & :j\neq k \\
0 & :j=k \end{array}\right.
\]
where
\[
S_{i,j,k} = 1[\{i\}]\cdot (B^{(i-1)}_{1,j+2}\cdot B^{(i-1)}_{2,k+2}+B^{(i-1)}_{1,k+2}\cdot B^{(i-1)}_{2,j+2})
\]
The squeeze factor of the stage is defined by
\[
\beta^{(i)}=1[\emptyset] + 1[\{i\}]\cdot B^{(i-1)}_{1,2}
\]

\subsection{Correctness Analysis.}
\label{sec: alg}
The essence of our algorithm is captured in the following lemma:
\begin{lem}
\label{lem: haf}
For every subset $X\subseteq [i-1]$
\[(\operatorname{haf}(B^{(i-1)}))_{X}=(\beta^{(i)}\cdot \operatorname{haf}(B^{(i)}))_{\{i\}\cup X}\]
for all $i\geq 1$.
\end{lem}

\begin{pf}
Recall that $B^{(i-1)}\in R[U_{i-1}]^{m \times m}$ with $m=2n-2i+2$.
By definition of  the hafnian
\begin{equation}
\label{eq: haf}
\mbox{haf}(B^{(i-1)}) =\sum_{\sigma\in C_{m}} \prod_{j=1}^{m/2} B^{(i-1)}_{\sigma(2j-1),\sigma(2j)}
\end{equation}
First consider the case $i=n$ separately. The above formula yields 
$\mbox{haf}(B^{(n-1)})=B^{(n-1)}_{1,2}$. We can write this as
\[
(\mbox{haf}(B^{(n-1)}))_\emptyset=((1[\emptyset]+1[\{n\}]\cdot B^{(n-1)}_{1,2})\cdot \mbox{haf}(B^{(n)}))_{\{1\}}
\]
since $\mbox{haf}(B^{(n)})=1$ by definition. Identifying the right hand side parantheses expression as $\beta^{(n)}$, it follows that the lemma is true for $i=n$.

Next consider the case $i<n$.
First observe that for every $\sigma\in C_m$,  we always have $\sigma(1)=1$ and either $\sigma(2)=2$ or $\sigma(3)=2$. We separate these two cases in Eq~\ref{eq: haf}:

\[
\mbox{haf}(B^{(i-1)}) =\underbrace{B^{(i-1)}_{1,2}\cdot \sum_{\substack{\sigma\in C_{m}\\ \sigma(2)=2}} \prod_{j=2}^{m/2} B^{(i-1)}_{\sigma(2j-1),\sigma(2j)}}_{T1}+\\
 \underbrace{\sum_{\substack{\sigma\in C_{m}\\ \sigma(3)=2}} B^{(i-1)}_{1,\sigma(2)}B^{(i-1)}_{2,\sigma(4)}\prod_{j=3}^{m/2} B^{(i-1)}_{\sigma(2j-1),\sigma(2j)}}_{T2}
\]
Second we note that the first term $T1$ of the right hand side above for every $X\subseteq [i-1]$
\[
T1_{X}=
\left(\!1[\{i\}]\cdot B^{(i-1)}_{1,2}\cdot\!\!\!\! \sum_{\sigma\in C_{m-2}}\!\!\prod_{j=1}^{m/2-1} \!\!\!B^{(i)}_{\sigma(2j-1),\sigma(2j)}
\!\right)_{\{i\}\cup X}
\] 
Third we see that the second term $T2$ of the right hand side for every $X\subseteq [i-1]$
\[
 T2_{X}
 =\left( \sum_{\sigma\in C_{m-2}}\prod_{j=1}^{m/2-1} B^{(i)}_{\sigma(2j-1),\sigma(2j)}
\right)_{\{i\} \cup X}
\]
Putting the two together we get
\[
\left(\mbox{haf}(B^{(i-1)})\right)_{X}=\\
\left(\!(1[\emptyset]\!+\!1[\{i\}]\cdot B^{(i-1)}_{1,2})\cdot \!\!\!\!\!\sum_{\sigma\in C_{m-2}}\!\! \prod_{j=1}^{m/2-1} \!\!B^{(i)}_{\sigma(2j-1),\sigma(2j)} \!\right)_{\{i\} \cup X}
\]
By the definition of the hafnian and the stretch factor, the Lemma follows. \qed
\end{pf}
Note that $\mbox{haf}(B)=(\mbox{haf}(B^{(0)}))_{\emptyset}$. Recursive application of Lemma~\ref{lem: haf} up to $\mbox{haf}(B^{(n)})=1$ (since $B^{(n)}$ is zero-dimensional), shows that $\mbox{haf}(B)=(\prod_{i=1}^n \beta^{(i)})_{[n]}$, which is what Algorithm~\ref{alg: haf} computes.
\subsection{Runtime Analysis.}

It is straightforward to note that adding two elements in $R[U_m]$ can be done in $2^m$ additions over $R$. Multiplication of two elements $a,b\in R[U_m]$ can be computed in $m^22^m$ additions and multiplications over $R$ using fast subset convolution \cite{BHKK07}. We describe its components here anew for completeness.
 
First, we introduce a polynomial over $R$ in a symbolic rank variable $r$, and compute the ranked zeta transforms
\[
\hat{a}_X(r)=\sum_{Y\subseteq X} a_Yr^{[Y]},\,\, \hat{b}_X(r)=\sum_{Y\subseteq X} b_Yr^{[Y]}
\]
Second, we compute the elementwise products
\[
\hat{c}_X(r)=\hat{a}_X(r)\hat{b}_X(r)
\]
Third, we take the ranked M\"obius inversion. Here $[r^m]p(r)$ for a polynomial $p(r)$ refers to the coefficient of the monomial $r^m$ in $p(r)$:
\[
d_X=[r^{|X|}]\sum_{Y\subseteq X} (-1)^{m-|Y|} \hat{c}_X(r)
\]
This $d$ is the product. The proof is implicit in Lemma~\ref{lem: op} in the next section, where we show how to make the whole algorithm use only polynomial space. For now, since both the zeta transform and the M\"obius inversion can be computed in $O(m2^m)$ ring operations by Yates's algorithm~\cite{Y37}, the runtime bound for multiplication follows.

Now confer Algorithm~\ref{alg: haf}. In stage $i$, we compute $B^{(i)}$ using at most $O((2n-2i)^2)$ multiplications and additions of $R[U_i]$ elements. The total runtime in terms of ring operations over $R$ of computing all the $B_i$'s is $\sum_{i=1}^n O((2n-2i)^2)O(i^22^i)\subset O^*(2^n)$.
Moreover, multiplying all the squeeze factors together is 
$\sum_{i=1}^n O(i^22^i) \subset O^*(2^n)$
ring operations on $R$. Altogether, an $O^*(2^n)$ ring operations algorithm as claimed.

\subsection{Polynomial Space.}
\label{sec: polyspace}
Explicitly holding a ring element of $R[U_m]$ requires storing up to $2^m$ elements of $R$. We describe in this section how the algorithm can be implemented to only hold a polynomial in $n$ number of elements of $R$ at any instant in time. The idea goes back at least to Karp~\cite{K82} who showed how some dynamic programming recurrences across subsets could be replaced by an inclusion--exclusion formula. The technique was recently formalized by Lokshtanov and Nederlof~\cite{LN10}. We describe it here for our algorithm in particular for completeness.

First we note that Algorithm~\ref{alg: haf} can be viewed as an evaluation of an arithmetic circuit over $R[U_n]$. We never do anything else intermediately with an element of $R[U_m]$ anywhere in the algorithm except for addition or multiplication with another element. Only at the very end we ask for the element of $h$ corresponding to $[n]$. 

One way of looking at the technique, is to consider the ranked zeta transforms of the previously section, and count in the transformed domain "all-the-way".
We don't transform back and forth every time we need to multiply two ring elements of $R[U_n]$, instead we stay in the ranked zeta transformed representation during the whole algorithm.
In the end we transform back via an inclusion--exclusion formula, since we are only interested in the element for $[n]$.
Let us again recall the ranked zeta transform.
For $e\in R[U_n]$ associate for every subset $X\subseteq [n]$ a polynomial in a symbolic rank variable $r$
\[
\hat{e}_X(r)=\sum_{Y\subseteq X} e_Yr^{[Y]}
\]
The beauty of this relation is that both additions and multiplications over $R[U_n]$ now can be treated as elementwise operations. This is well known, see for instance \cite{BHKK07}.
The case of addition is straightforward from the definition, i.e. since $(e+f)_{X}=e_{X}+f_{X}$ we have $\widehat{e+f}_X=\hat{e}_X+\hat{f}_X$ from linearity.  For multiplication we have

\begin{lem}[\cite{BHKK07}]
\label{lem: op}
For any $e_1,e_2,\dots,e_k \in R[U_n]$
\[
(e_1 \cdot e_2 \cdots e_k)_{[n]}=
[r^n]\sum_{X\subseteq [n]} (-1)^{n-|X|}\widehat{(e_1\cdot e_2 \cdots e_k)}_X(r) =
[r^n]\sum_{X\subseteq [n]} (-1)^{n-|X|} \prod_{i=1}^k \hat{e_i}_X(r)
 \]
 \end{lem}
 \begin{pf}
We recall from the definition of multiplication for the first (left) expression that 
\[
(e_1\cdot e_2 \cdots e_k)_{[n]}=
\sum_{\substack{Y_1\cup Y_2 \cup \cdots \cup Y_k=[n]\\ \forall i<j:Y_i\cap Y_j=\emptyset}} \prod_{i=1}^k e_{iY_i}
\]
For the middle expression, we only have to plug in the definition of the ranked zeta transform to arrive at the same value
\[
[r^n]\!\sum_{X\subseteq [n]} (-1)^{n-|X|}\widehat{(e_1\cdots e_k)}_X(r) =
[r^n]\!\sum_{X\subseteq [n]} (-1)^{n-|X|}r^{|X]} \!\!\!\!\!\!\!\! \sum_{\substack{Y_1\cup \cdots \cup Y_k=X\\ \forall i<j:Y_i\cap Y_j=\emptyset}} \prod_{i=1}^k e_{iY_i} =
\!\!\!\!\!\!\!\!
\sum_{\substack{Y_1\cup \cdots \cup Y_k=[n]\\ \forall i<j:Y_i\cap Y_j=\emptyset}} \prod_{i=1}^k e_{iY_i}
\]
Finally, for the right expression
\[
[r^n]\sum_{X\subseteq [n]} (-1)^{n-|X|}\prod_{i=1}^k \hat{e_i}_X(r) = 
[r^n]\sum_{X\subseteq [n]} (-1)^{n-|X|}(\prod_{i=1}^k \sum_{Y_i\subseteq X} e_{iY_i}r^{|Y_i|}) =
\sum_{\substack{Y_1\cup Y_2 \cup \cdots \cup Y_k=[n]\\ \forall i<j:Y_i\cap Y_j=\emptyset}} \prod_{i=1}^k e_{iY_i}
\]
where the last expression follows after collecting all terms contributing to the coefficient of $r^n$, and noting that contributions for $Y_i$'s such that $\bigcup_i Y_i\subset [n]$ are counted equally many times with the sign $+$ as $-$ (since a nonempty set has as many even sized subsets as odd sized ones).
\qed
\end{pf}

Via Lemma~\ref{lem: op} and the linearity of addition, we see that the quantity Algorithm~\ref{alg: haf} outputs, $(\beta^{(0)}\cdot \beta^{(1)} \cdots \beta^{(n)})_{[n]}$, can be computed elementwise in the transformed domain. Thus we can lift out the summation across the subsets to the outermost level. Every arithmetic operation can be carried out in the transformed domain over a degree $n$ polynomial in $R[r]$, module $r^{n+1}$ since we only care about the coefficient of $r^n$. 
Our new polynomial space algorithm is given in Algorithm~\ref{alg: polyspace}. Here, the matrices $\hat{B}^{(i)}\in R[r]^{(2n-2i)\times (2n-2i)}$, and the squeeze factors $\hat{\beta}^{(i)}\in R[r]$ are transformed analogues of $B^{(i)}$ and $\beta^{(i)}$ from Algorithm~\ref{alg: haf}. Note that the increase in runtime is at most polynomial in $n$.

\begin{algorithm}
\caption{\label{alg: polyspace}
haf($B$),$\,B\in R^{2n\times 2n}$}
\begin{algorithmic}
\STATE $h\leftarrow 0$
\STATE $\hat{B}^{(0)}\leftarrow B$
\FOR{$X\subseteq [n]$}
\STATE $g\leftarrow 1$
\FOR{$i=1\cdots n$}
\IF{$i\in X$}
\STATE $\hat{B}^{(i)}_{j,k}=r(\hat{B}^{(i-1)}_{1,j+2}\hat{B}^{(i-1)}_{2,k+2}+\hat{B}^{(i-1)}_{1,k+2}\hat{B}^{(i-1)}_{2,j+2})+\hat{B}^{(i-1)}_{j+2,k+2}:j\neq k, 0:j=k.$ 
\STATE $\hat{\beta}^{(i)}=1+r\hat{B}^{(i-1)}_{1,2}$ 
\ELSE
\STATE  $\hat{B}^{(i)}_{j,k}=\hat{B}^{(i-1)}_{j+2,k+2}$ 
\STATE $\hat{\beta}^{(i)}=1$
\ENDIF
\STATE $g\leftarrow g\hat{\beta}^{(i)}$
\ENDFOR
\STATE $h\leftarrow h+(-1)^{n-|X|}[r^n]g$
\ENDFOR
\RETURN $h$
\end{algorithmic}
\end{algorithm}


\section{The Hardness of General Exact Set Cover Counting }
\label{sec: hard}
In this section we prove Theorem~\ref{thm: hard}.
A general exact set cover problem consists of a ground set $U$ on $n$ elements, and a family $\mathcal{F}$ of subsets of $U$. The objective is to find a subset $\mathcal{F}'\subseteq \mathcal{F}$ such that I) $\cup_{F\in \mathcal{F}'} F=U$ and II) $\forall F\neq G\in \mathcal{F}':F\cap G=\emptyset$. In the counting analogue we are asked to count the number of such covers.
This extends the problem of counting perfect matchings in a graph $G=(V,E)$ since $U=V$ and $\mathcal{F}=E$ is a valid instance. One might suspect that the algorithm presented in this paper after some careful tailoring would improve the runtime of general exact set cover counting. Unfortunately this seems not to be the case, at least not significantly. We show here that if it did, we would have faster algorithms for the $0/1$ matrix permanent.

Let $\epsilon_1>0$ be a fixed constant less than one. For large enough integers $k,n$ we can have
$n$ chosen as the largest integer smaller than $2^{\lceil (\epsilon_1k/2-1)\rceil}$ for which $k$ divides $n$. Note that for such values $k\geq \epsilon_1^{-1}(2\lceil \log_2 n\rceil+2)$. 
Let $M\in \{0,1\}^{n\times n}$ be a matrix whose permanent we want to compute.
First note that the permanent of $M$ is a positive integer at most equal to $n!$. 
We will use the Chinese remainder theorem to recover the value of the permanent.
We compute $a_i=\mbox{per}(M) (\mbox{mod } m_i)$ for $n$ pairwise coprime moduli $m_i < n^2$. The prime number 
theorem asserts that we can find such a set of moduli for large enough $n$. Just take $m_i$ as the smallest power larger than $n$ of the $i$:th prime.

We show how to construct counting exact set cover instances $\mathcal{I}_i=(U,\mathcal{F}_i)$ for each $m_i$, whose solution equals $a_i(\mbox{mod }m_i)$. We start by dividing the rows of $M$ into $m$ groups $R_1,R_2,\cdots, R_m$ of the same size $k$. 
We will add sets to $\mathcal{F}_i$ for each such group so that in every exact set cover, exactly one set from each group is present. To this end, we associate a set $L_i$ for each group $R_i$, which will all be part of the ground set $U$.
Each $L_i$ has size $2\lceil \log_2 n\rceil+2\leq \epsilon_1k$. We let $l\neq r\in L_i$ be two of its elements.
The subsets of $L_i$ containing exactly one of $l$ or $r$, are divided in two families $\mathcal{L}_i$ and $\mathcal{R}_i$ as follows. Every subset $T\subseteq L_i-\{r\}$ such that $l\in T$ belongs to $\mathcal{L}_i$, and every subset $T\subseteq L_i-\{l\}$ such that $r\in T$ belongs to $\mathcal{R}_i$.
Note that for any two disjoint  $T_1,T_2\in \mathcal{L}_i \cup \mathcal{R}_i$ such that $T_1 \cup T_2=L_i$, one of them belong to $\mathcal{L}_i$ and the other to  $\mathcal{R}_i$.
Also note that $\mathcal{L}_i$ has at least $n^2>m_i$ members.

The construction is as follows. We set $U=[n] \cup L_1 \cup \cdots \cup L_k$. Note that $|U|\leq n + \epsilon_1n$.
Let $\mathcal{U}_k$ denote the family of all $k$-sized subsets of $[n]$.
For every subset $T\in \mathcal{U}_k$ and $j$ we compute
\[
w_j(T)= \sum_{\sigma \in S_k} \prod_{l=1}^k M(R_j(l),T(\sigma(l))) \,\,(\mbox{mod } m_i)
\]
Here and in the following $T(l)$ for a set $T$ refers to the $l$th member of the set.
We let $\mathcal{F}_i$ contain the sets  $T\cup \mathcal{L}_j(1),T\cup \mathcal{L}_j(2),\cdots, T\cup \mathcal{L}_j(w_j(T))$ for every $T\in \mathcal{U}_k$ and every $j$. We also add all sets in $\mathcal{R}_j$ for all $j$ to $\mathcal{F}_i$. Note that $|\mathcal{F}_i|\in O(|U|^{\log |U|})$.
This completes the construction.

\begin{lem}
\label{lem: Iper}
The solution to $\mathcal{I}_i$ modulo $m_i$ equals $a_i$.
\end{lem}
\begin{pf}
By definition
\[
\mbox{per}(M)=\sum_{\sigma\in S_n} \prod_{i=1}^n M_{i,\sigma(i)}
\]
Identifying the $m$ row groups the relation can be rewritten
\[
\mbox{per}(M)=\!\!\!\!\!\sum_{\substack{T_1\cup T_2 \cup \cdots \cup T_m=[n]\\ \forall a: T_a\in \mathcal{U}_k}} \prod_{j=1}^m \sum_{\sigma\in S_k} \prod_{i=1}^k M(R_j(i),T_j(\sigma(i)))
\]
Substituting $w$ into the formula we have
\begin{equation}
\label{eq: perid}
\mbox{per}(M) \equiv \sum_{\substack{T_1\cup T_2 \cup \cdots \cup T_m=[n]\\ \forall a: T_a\in \mathcal{U}_k}} \prod_{j=1}^m w_j(T_j) \,\,(\mbox{mod } m_i)
\end{equation}
From the other direction, note that every exact set cover $\mathcal{F}'\subset \mathcal{F}_i$ contains exactly $2m$ sets, one set from each $\mathcal{R}_j$, and for every $j$ there is a set which is the union of a set  $T$ from $\mathcal{U}_k$ and one of the first $w_j(T)$ sets in $\mathcal{L}_j$. Let $\mathcal{L}_j[l]$ be shorthand for the first $l$ sets in $\mathcal{L}_j$, i.e. the sets $\mathcal{L}_j(1),\mathcal{L}_j(2),\cdots,\mathcal{L}_j(l)$. We have that the solution to $\mathcal{I}_i$ equals
\[
\sum_{\substack{V_1\cup W_1 \cup V_2 \cup W_2\cdots \cup W_{2m}=U\\ \forall j: V_j\cap L_j\in \mathcal{L}_j[w_j(V_j\cap[n])] \\ \forall j: V_j\cap [n]\in \mathcal{U}_k \\ \forall j: W_j \in \mathcal{R}_j}} 1
\]
Since the $W_j$'s are uniquely determined by the $V_j$'s,  it can be rewritten as
\[
\sum_{\substack{V_1\cup V_2 \cup \cdots \cup V_m=U\\ \forall j: V_j\cap L_j \in \mathcal{L}_j[w_j(V_j\cap [n])] \\ \forall j: V_j \cap [n]\in \mathcal{U}_k}} \!\!\!\!\!\!\!\!\!\! 1 \equiv \sum_{\substack{T_1\cup \cdots \cup T_m=[n]\\ \forall a: T_a\in \mathcal{U}_k }} \prod_{j=1}^m  w_j(T_j) \,\,(\mbox{mod } m_i)
\]
This is the same expression as the right hand side of Eq.~\ref{eq: perid}.
\qed
\end{pf}

We set $\epsilon_2=\epsilon_1^2$. It takes $O(m\binom{n}{k}k!)$ time to construct each $\mathcal{I}_i$ na\"ively from $M$. For our choice of $k$, this is (much) less than $O^*(2^{(1-\epsilon_2)n})$ for large enough $n$. Now if we had an $O^*(2^{(1-\epsilon_1)n'})$ time algorithm to solve an exact set cover counting problem on an $n'$ element ground set, we could solve $\mathcal{I}_i$ for all $i=1,2,\cdots,n$ and hence compute $\mbox{per}(M)$ from the $a_i$'s using the Chinese remainder theorem in $O^*(2^{(1-\epsilon_1^2)n})=O^*(2^{(1-\epsilon_2)n})$ time, since $n'=|U|=(1+\epsilon_1)n$.
This completes the proof of Theorem~\ref{thm: hard}.

\section*{Acknowledgments}
The author is grateful to several anonymous referees for valuable comments.

\end{document}